\documentclass[DIV12]{scrartcl}
\usepackage{graphicx}
\usepackage{amsmath}
\usepackage{siunitx}
\usepackage{authblk}
\usepackage{cite}
\usepackage{color}
\usepackage{ulem}
\usepackage[latin1]{inputenc}
\setkomafont{disposition}{\normalcolor\rmfamily}
\setkomafont{caption}{\small\mdseries\selectfont}

\setcapindent{0mm}
\DeclareMathAlphabet{\mathcal}{OMS}{cmsy}{m}{n}
\newcommand{\removed}[1] {}

\newcommand{\indur}[2]{#1_{\text{#2}}}
\newcommand{\cbs} {CBS}
\newcommand{\sbn} {SBN}
\newcommand{\nrp} {NRP}
\newcommand{\gpb} {GPB}
\newcommand{\pgs} {\sigma}
\newcommand{\cax}{$c$-axis}
\newcommand{\fig}[1]{Fig. \ref{#1}}

\newcommand{\kt}{\indur{k}{t}}
\newcommand{\vkin}{\indur{\vec{k}}{in}}
\newcommand{\kin}{\indur{k}{in}}
\newcommand{\vkout}{\indur{\vec{k}}{out}}

\newcommand{\vkcbs}{\indur{\vec{k}}{\cbs{}}}

\begin{document}
\title{\huge Observation of transverse coherent backscattering in disordered photonic structures}

\author{\large Sebastian Brake$^{1}$, Martin Boguslawski$^{1,*}$, Daniel Leykam$^{2}$, Anton Desyatnikov$^{2}$, and Cornelia Denz$^{1}$}

\affil{\normalsize
$^1$Institut f\"ur Angewandte Physik and Center for Nonlinear Science (CeNoS), \\Westf\"alische Wilhelms-Universit\"at M\"unster, 48149 M\"unster, Germany\\
$^2$Nonlinear Physics Centre, Research School of Physics and Engineering, The Australian National University, Canberra ACT 0200, Australia
\\
$^*$Corresponding author: martin.boguslawski@uni-muenster.de
}
\date{}
\maketitle
\begin{abstract}
We report on the experimental observation of weak localization in an optically induced disordered (2+1)-dimensional photonic structure. Our flexible method of optical induction is applied with a nondiffracting random intensity distribution. We focus on the analysis of a statistical output spectrum for many probe events with variance of the incoming beam's transverse spatial frequency. For particular spatial frequencies we find considerable signatures of transverse coherent backscattering.
\end{abstract}

Coherent backscattering (\cbs{}) is a pure wave effect evolving from scattering in random media.
Its appearance can briefly be described as incoming wave vectors are in average reversed such that the probability to find outcoming waves in the antiparallel direction of incidence is enhanced. \cite{Bergmann}
This effect bases upon wave interference and, given that a random medium has appropriate conditions, it can be found among various systems where sufficient long wave propagation is provided.
Thus, numerous experiments were designed and realized in order to explore the conditions yielding to \cbs{}, and to prove the universality of this effect, e.g. by applying acoustic \cite{Kirkpatrick, Bayer, Tourin} and seismic \cite{Larose} as well as matter \cite{Stellmach, Jendrzejewski} and light waves in turbid colloids \cite{vAlbada, Wolf} as well as in cold atoms \cite{Labeyrie1999, Bidel}.
Aiming to find a wave which scatters preferably elastically, especially by using light as the preferred wave, a lot of efforts have been invested during the last three decades in order to clarify scattering processes inside a turbid medium \cite{Wiersma, Gross}---a comprehensive review can be found in Ref. \cite{Aegerter}, and to establish a universal theory which is in consistence with all experimental results \cite{Skipetrov2004, Skipetrov2006}.
Since the scaling theory predicts that there is a transition from extended to localized states in a system of more than two dimensions, \cite{Abrahams} special attention has been payed to describe localization in three-dimensional samples. \cite{Aegerter}
However, from experiments on three-dimensional random media it is still hard to distinguish whether reduced transmission or reflection rates come from localization or from absorbing and resonant processes. \cite{Berkovits, vTiggelen}
Still, especially identifying and examining the critical regime between localizing and diffusive behavior is a burning issue to be answered. \cite{Sperling}

Though all modes are localized in one- and two-dimensional random systems, media of these dimensionalities are of high interest to investigate localization and backscattering events in various configurations as they allow for a deep insight into these wave events in a less complex manner. \cite{Jendrzejewski, Labeyrie2012}
For there is a distinction between strong and weak localization (commonly referred to Anderson localization \cite{Anderson} and Coherent Backscattering), it is moreover most important to find model systems in which can be switched between both regimes and additionally compared to ballistic propagation at will.
It has been shown in several publications \cite{Schwartz, Levi2011, Levi2012, Boguslawski} that in an optically induced randomized potential ensemble the observation of transverse strong localization is achievable by defining proper conditions.

In this contribution we describe the preparation of proper ensembles of two-dimensional random photonic potential landscapes to experimentally observe and investigate \cbs{} signatures.
The potentials in terms of disordered refractive index modulations are optically induced in a photorefractive strontium barium niobate crystal (\sbn{}).

\begin{figure}[tbh]
  \center
	\includegraphics[width=.4\textwidth]{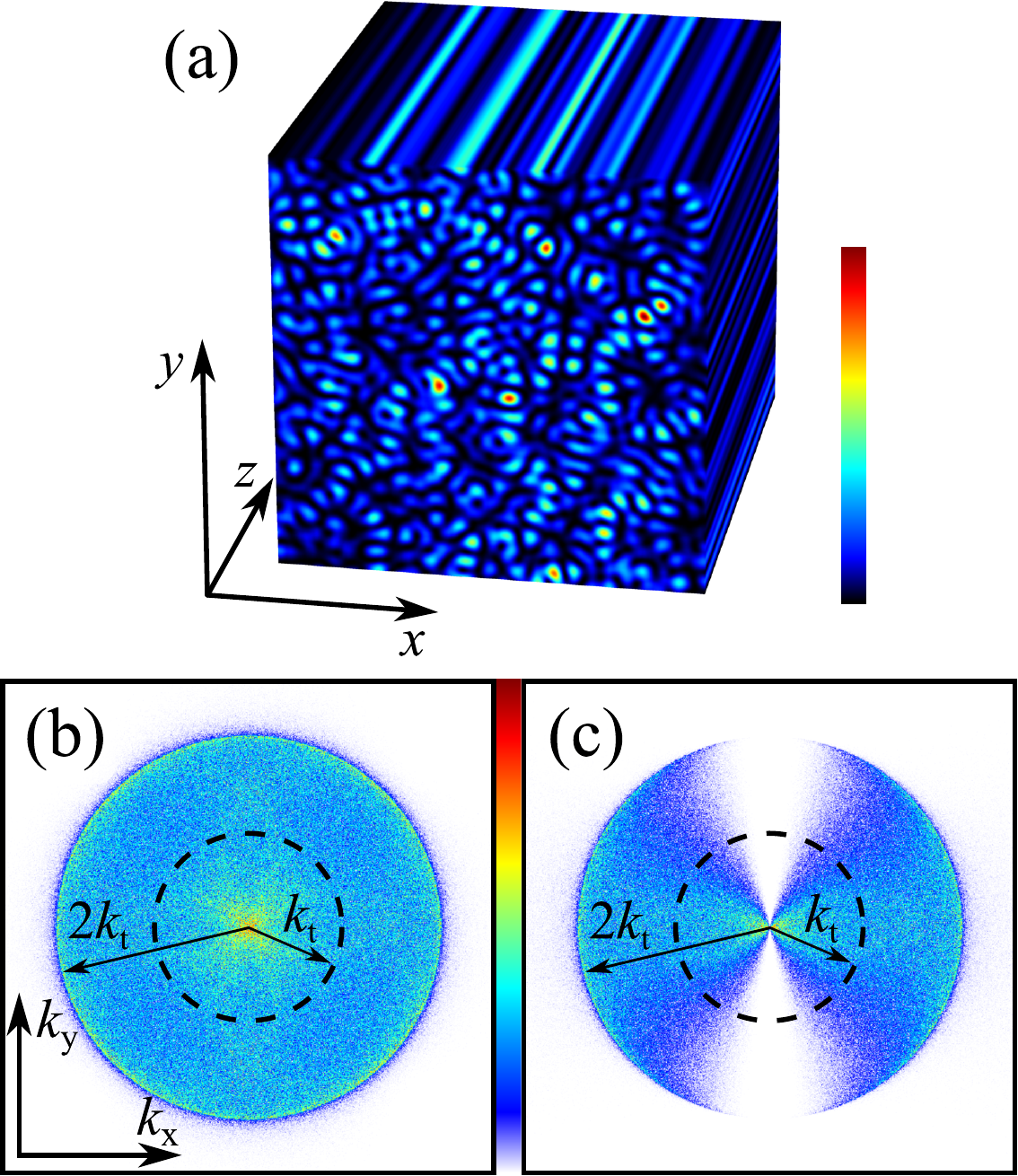}
	\caption{
	(a) Exemplifies a three-dimensional plot of volume of intensity for a single nondiffracting random intensity distribution. Modulation of intensity only occurs in $x$-$y$ plane and agrees in every plane along direction of propagation $z$. (b) Power spectrum of effective intensity resulting from incoherent superposition of 100 different \nrp{}s of spectral radius $\kt{}$ (position marked by dashed circle). (c) Power spectrum of corresponding averaged refractive index distribution originating from effective intensity.
	}
	\label{fig:propPotential}
\end{figure}

In order to induce such a configuration we employ so-called nondiffracting random intensity patterns (\nrp{}) \cite{Cottrell} with the transverse intensity being randomly distributed.
To get an impression of these intensity distributions, a three-dimensional volume is scetched in \fig{fig:propPotential} (a) where the trihedron clarifies the orientation and direction of propagation which is congruent with $z$.
As for all nondiffracting beams, this distribution is constant along a particular propagation distance, i.e. the intensity is independent of the third dimension, according to the direction of propagation. \cite{Durnin}
Selecting appropriate parameters for the structural size of such intensity---for these random beams we introduced the nomenclature photonic grain size $\pgs{}$ in an earlier publication \cite{Boguslawski}---one can reach nondiffracting distances of several centimeters and even longer.
Throughout all the experiments presented here, we fix $\pgs{}$ to \SI{20}{\micro\meter} which is connected to a nondiffracting distance that is longer than the long site of the \sbn{} crystal of dimension \SI{5 x 20 x 5}{\milli\meter\cubic}.
The nondiffracting property makes these light configurations particularly appropriate to implement (2+1) dimensional systems and thus to mimic the temporal development in a two-dimensional photonic configuration. \cite{Weilnau}

With these ingredients to form (2+1)-dimensional, transversely disordered potentials we establish an adequate platform to examine weak transverse localization, involving coherent backscattering events in transverse direction, i.e. perpendicular to the direction of light propagation.
It has already been shown that regimes of strong localization can be created in order to present transverse Anderson localization in disordered regular but also completely randomized photonic lattice systems. \cite{Schwartz, Levi2011, Boguslawski}

For the induction process of one single potential configuration we supply an external field of a few \si{\kilo\volt\per\centi\meter} and a writing light field whose intensity contains the structure of the desired potential and which is ordinarily polarized with respect to the crystal's symmetry axis (\cax{}).
During the induction process charge carriers of the \sbn{} crystal rearrange according to the intensity modulation and cause internal electrical fields which in turn induce a refractive index modulation due to the linear electro-optic effect \cite{Hall}.
Due to the (2+1)-dimensional character of the potential, the induced refractive index distribution can be taken as an arrangement of waveguides where randomization is introduced with respect to the waveguide depth and position.
Accordingly, every waveguide is potentially a scattering center with the ability to change the wave vector $\vec k$ of an incoming light field.

The character of \cbs{} can be analyzed best when averaging over a large ensemble of random potentials.
Introducing an appropriate ensemble, we experimentally use 200 various \nrp{} configurations, for the induction of each potential we apply an external field of \SI{2}{\kilo\volt\per\centi\meter} and set the illumination time of an individual writing beam to \SI{10}{\second}.
In \fig{fig:propPotential} (b), the numerically computed power spectrum of an effective intensity resulting from the average of 100 transverse \nrp{} distributions is presented.
Circular spectra of all contributing \nrp{} have the spectral radius $\kin{}$, whereas the limiting spatial frequency for the power spectrum is $2\kin{}$.
This can be understood by simple mathematic arguments as the power spectrum is the convolution of the \nrp{} spectrum of radius $\kt{}$ with itself.
As the spectrum of a \nrp{} is a circle, the power spectrum of the effective intensity accordingly is a filled disk of radius $2 \kt{}$.
For comparison to the mean intensity spectrum, a numerical calculation of the power spectrum resulting from the average of all random potentials is depicted in \fig{fig:propPotential} (c).
Due to an orientation anisotropy in \sbn{} crystals,\cite{Terhalle} the power spectrum of this mean refractive index modulation gets deformed as the modulation perpendicular to the \cax{} drops to zero.
This characteristic is prominent in \fig{fig:propPotential} (c) where many scattering centers can be found in $\indur{\vec{k}}{x}$ direction in contrast to $\indur{\vec{k}}{y}$ showing no modulation just along this direction.
By reasons of this anisotropy, we limit our experiments to the investigation of scattering characteristics in direction of $\indur{\vec{k}}{x}$.

The probing process of each potential is performed in the linear propagation regime (cw laser light of a feq \si{\micro\watt}s at a wavelength of $\lambda=$ \SI{532}{\nano\meter}) with a broad Gaussian probe beam (\gpb{}), polarized extraordinarily to the \cax{}.
Holding a fixed Gaussian beam waist of $w_0 = $ \SI{200}{\micro\meter} that covers several scattering centers at the input facet, we introduce a variable tilt of the probe beam by a nonzero transverse component $\vkin{}$ in order to address various spatial input frequencies.
All beams to optically induce the potential as well as to probe the induced potential are generated by use of a set of spatial light modulators \cite{Boguslawski}.
Single output spectra were recorded by a CCD camera placed in the Fourier space of the \sbn{}'s output facet.
We further subsequently average all output spectra of induced potential realizations receiving a spectrum of probability $P(\vkin{})$ to identify most probable spatial output frequencies.
To do so experimentally, we benefit from the reversibility property of the SBN crystal since after probing one potential configuration it can be erased by white-light LED arrays in order to induce another structure afterwards.
After typical erasure times of \num{30} to \SI{40}{\second}, the refractive index distribution is homogenized again.

\begin{figure}[t]
  \center
	\includegraphics[width=.5\textwidth]{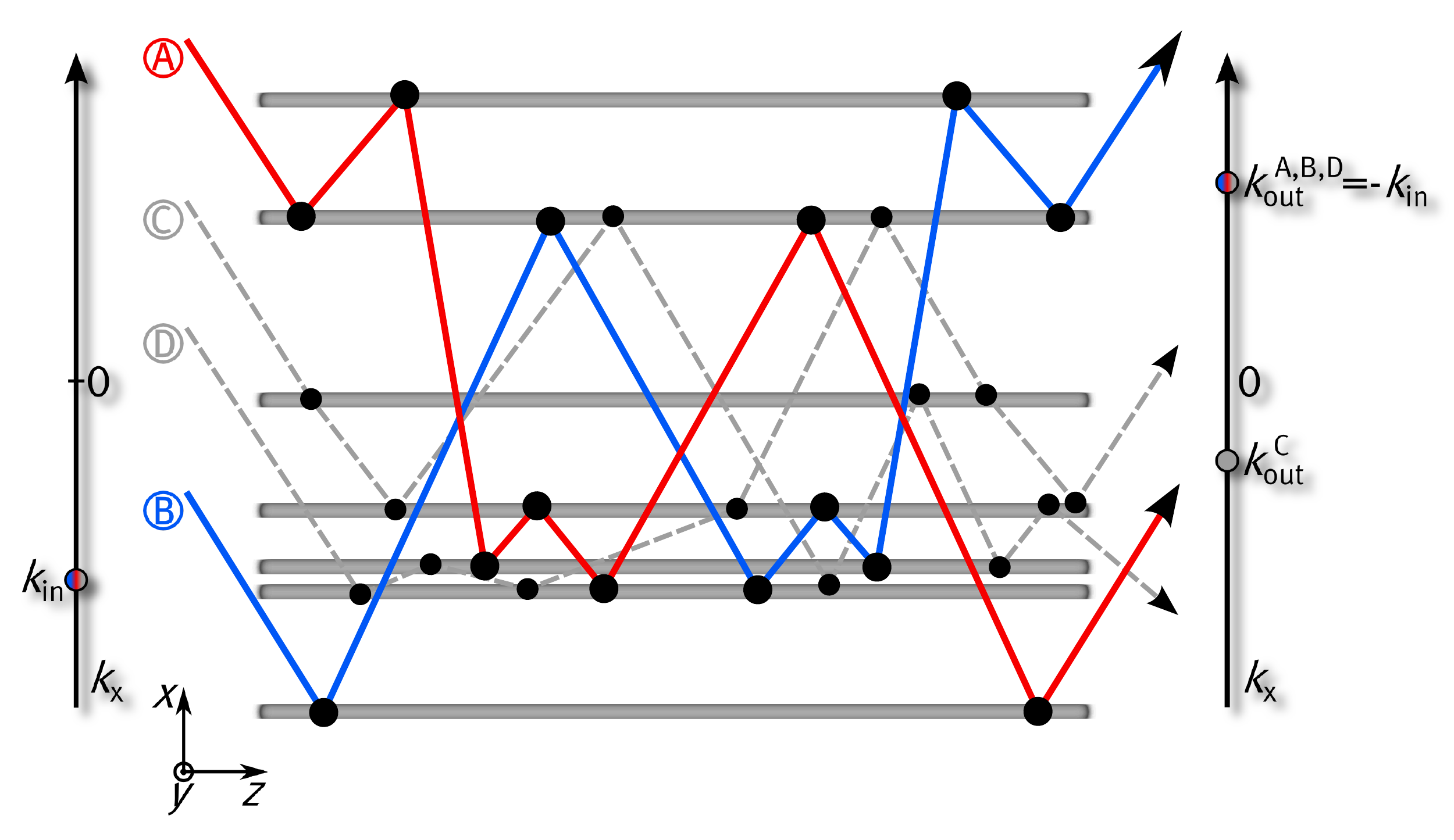}
	\caption{
	Schematic of waveguides of the described (2+1)-dimensional photonic system of disorder. Gray lines represent projection of waveguides to the $x$-$z$ plane. Lines are possible light paths scattering through waveguide arrangement. Blue path $B$ is reversion of $A$ (red), only those path pairs contribute to coherent backscattering. Gray paths symbolize arbitrary light paths with random phase difference to $A$ or $B$.
	}
	\label{fig:wgSchema}
\end{figure}

In \fig{fig:wgSchema} a schematic illustrates the projection to the $x$-$z$ plane of one of the considered (2+1)-d potential realizations where scattering centers are represented by gray straight bars.
Four possible paths of light are indicated propagating from the input to the output facet of the potential, each with the same input wave vector $\vkin{}$ and individual output vectors $\vkout{}$.
Path $B$ thereby is the time inverted path of $A$, thus, $A$ and $B$ have the same lengths. 
Additionally, $\vkin{} = -\vkout{}$ is valid (which is also valid for path $D$, but the path length is different from $A$ or $B$).
Hence, light fields that are propagating along $A$ and $B$ or along according path pairs interfere always constructively in the far field, since their transverse output frequency component is the negative of the input component.
In contrast to this, light that propagates along other paths may interfere constructively, destructively or, in general, can have any modulus of phase difference between $0$ and $2\pi$.
In consequence, these interferences average out after considering many potential realizations.
Solely, path pairs such as $A$ and $B$ will contribute constructively to the averaged output spectrum, namely exactly at the spatial frequency $\vkout{} = -\vkin{}$.

\begin{figure}[t]
  \center
	\includegraphics[width=.4\textwidth]{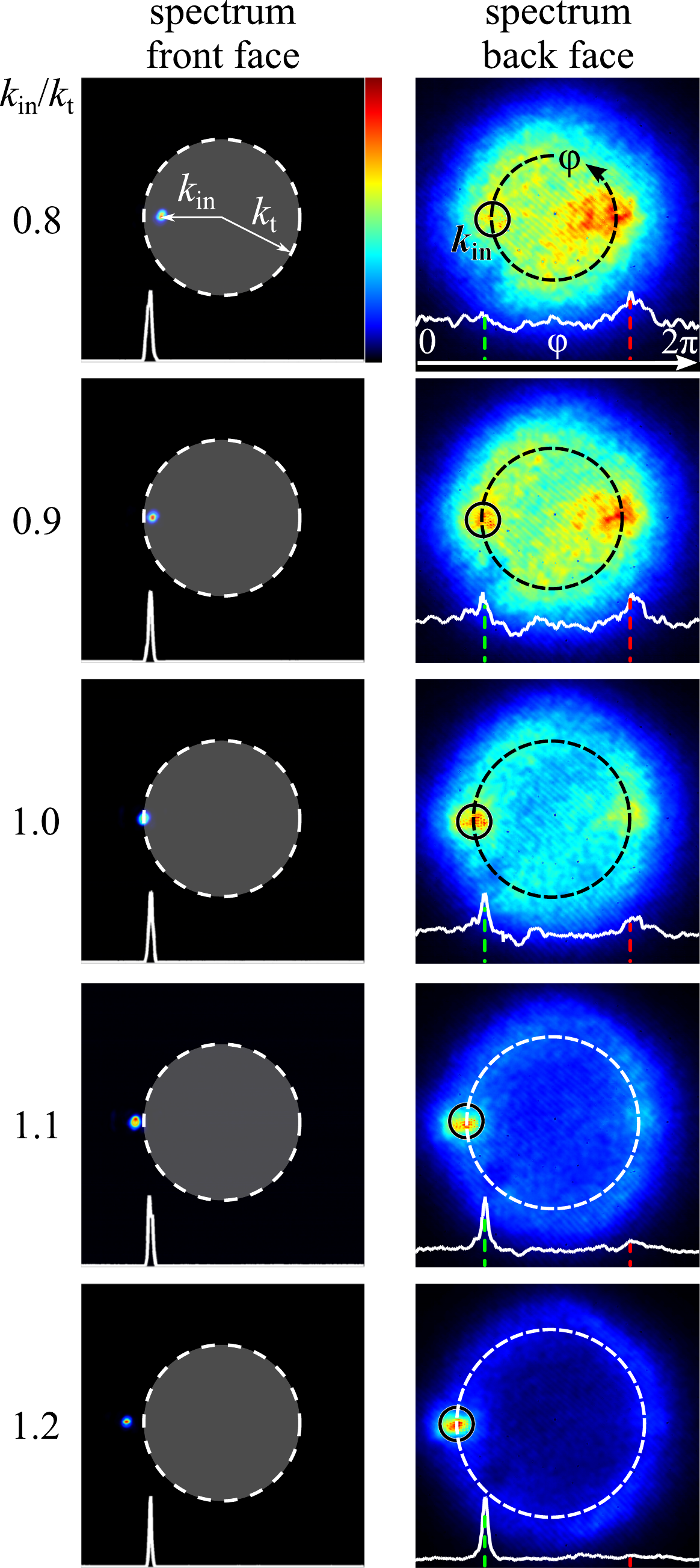}
	\caption{
	Upper row shows input spectra of \gpb{}s at various spatial frequencies $\kin{} /\kt{}$. Gray circles illustrate spectrum of \nrp{}. Lower row shows probability spectra corresponding to \gpb{}s of input frequencies shown in upper row. Plots at lower part indicate azimuthal spectral distribution $P(\varphi)$ at radius $\kin{}$. Green and red dashed lines mark $\vkin{}$ and $\vkcbs{} = -\vkin{}$.
	}
	\label{fig:expSpectra}
\end{figure}

The left column in \fig{fig:expSpectra} shows the input spectra of \gpb{}s of various tilts from $\kin{}/\kt{} = 0.8$ to $1.2$.
For comparison, gray dashed circles illustrate the input spectrum of the potential-writing \nrp{}s with fixed transverse wave vector component of length $\kt{} = \pi/\pgs{}$.
Corresponding probability spectra after probing all realizations of the disordered potential are shown in the right column.
Plots appended in the lower part of each image represent the azimuthal spectral distribution $P(\varphi)$ at radius $\kin{}$, where the angle $\varphi$ equals zero at 12 o'clock and increases counterclockwise (cf. upper right image in \fig{fig:expSpectra}).
In the right-column plots, input frequency $\vkin{}$ (dashed green line) and expected \cbs{} frequency $\vkcbs{}$ (dashed red line) are marked.
Following the spectra in the right column from top to bottom, we find that \cbs{} signatures are present for all tilts of \gpb{} as the intensity is increased around the spatial frequencies $\vkcbs{}$ where the \cbs{} peak is expected, respectively.
However, especially for a \gpb{} tilt of $\kin{} = 0.9\kt{}$, we clearly see indications of \cbs{} since for this parameter the $\vkin{}$ component of the mean output spectrum has the most significant contribution around the spectral position $\vkcbs{}$.

\begin{figure}[tbh]
  \center
	\includegraphics[width=.45\textwidth]{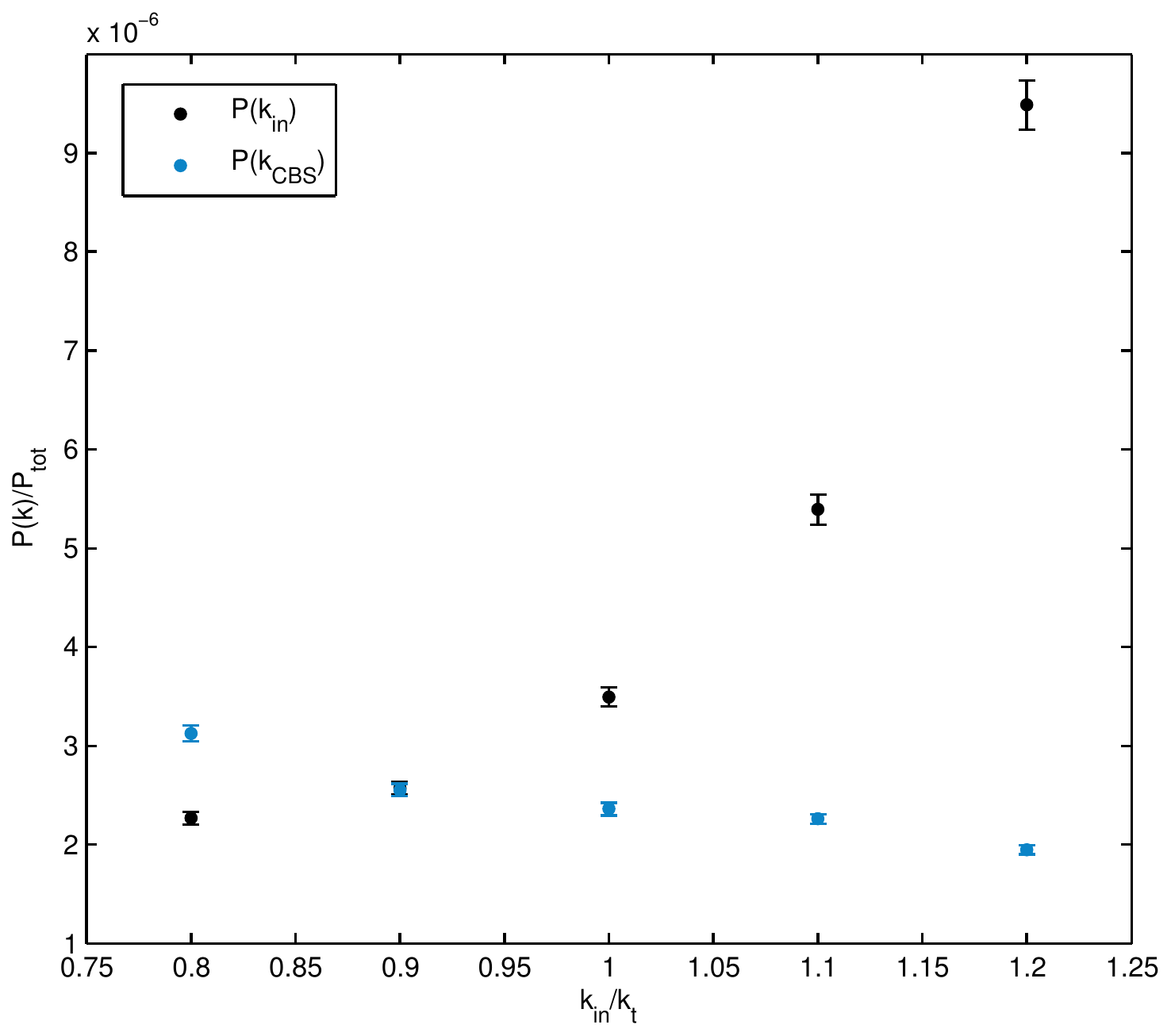}
	\caption{
	Relative power $P(k)$ of input frequency $\vkin{}$ (black) and expected \cbs{} frequency $\vkcbs{}$ (blue). Spatial frequency normalized to $\kin{}$, power normalized to total power of spectrum $P_{tot}$. Errors are estimated by standard error of the mean power of spatial frequency.
	}
	\label{fig:Pin_vs_Pcbs}
\end{figure}

To analyze this behavior more quantitatively, relative powers of particular spatial frequencies $P(k)$ are plotted against varied tilts in \fig{fig:Pin_vs_Pcbs}.
Here $P(k)$ is normalized to the overall power of the respective spectrum.
For small tilts around $\kin{}/\kt{} = 0.8$ the \cbs{} peak $P(\vkcbs{})$ (blue data points) is higher than for the input frequency $P(\vkin{})$ (black data points).
This behavior changes with increasing tilt as for high values of $\kin{}/\kt{}$ there is almost no light backscattered to the specific expected \cbs{} frequency.
It is obvious that the scattering probability is weakened for higher spatial probing frequencies, as a considerable percentage of light is transmitted rather than scattered (cf. \fig{fig:expSpectra} spectrum for $\kin{}/\kt{} = 1.2$).

Thus, since scattering processes are less probable for higher spatial probing frequencies, it seems that the propagation distance of \SI{20}{\milli\meter} is not sufficiently long for a prominent \cbs{} peak to be established.
For these short distances the probability for light to find proper paths are rather small, however increase with longer distances.
Consequently, by changing the tilt of our probing beam we can adjust the scattering strength of our potential.

Worth mentioning is the texture of the probability spectra apart from $\vkin{}$ and $\vkcbs{}$.
Finding an almost filled spectrum for small tilts, the spectrum for higher $\kin{}$ values becomes more and more ring-like.
The radius of this ring equals the spatial input frequency and spectral parts for small spatial frequencies become rather weak.
Hence, for higher tilts the transverse wave vector component is almost conserved but merely changed in orientation in contrast to flat-angled probing below $\kin{}$.
For these cases, a higher percentage of light is scattered to the spectral center.
Besides the direction, also the length of the transverse wave vector is changed.

Elastic scattering processes can be described by applying the Ewald construction to our mean power spectrum of the potential that holds a limit frequency of $2\kt{}$ (cf. \fig{fig:propPotential}).
Thereby all possible elastic scattering centers can be found on a circle of radius $\kin{}$ (the so called Ewald circle in two dimensions), touching the power spectral center which, together with the center of the Ewald circle, resembles $\vkin{}$.
The amplitude along the latter circle thereby determines the scattering strength.
Now it becomes obvious why scattering is weaker for higher $\kin{}$---the Ewald circle includes decresingly strong scattering centers as $\kin{}$ increases (cf. \fig{fig:propPotential}, (c)).
Thus at $2\kt{}$ our potentials reveal an effective mobility edge, where, approaching this edge, scattering becomes more and more improbable. \cite{Sanchez, Gurevich}

However, coherent processes such as \cbs{} cannot be explained with this model since the influence on the phase is neglected.
In this connection, it is also not allegeable how processes that lead to a change of the transverse wave number $\kin{}$ (not only the direction) happen.
Most likely, these processes go back to the influence of waveguiding for it is more probable that light is guided for a nonnegligible distance along local refractive index maxima before coupled to neighboring waveguides.
This scheme is supposed to change the transverse spatial frequency markedly.

In conclusion, we presented an optically induced disordered (2+1)-dimensional photonic system as a platform for the investigation of transverse coherent backscattering effects.
The presented results reveal that our approach is explicitly appropriate to introduce potential ensembles of variable disorder strength where both regimes, strong and weak scattering, can be examined.
We demonstrated that besides Anderson localization also \cbs{} is observable.
However, a few open points to be discussed are still left.
The photonic grain size $\sigma{}$ and the angle of incidence $\kin{}$ turn out to be appropriate as control parameters for the strength of localization.
Though we previously analyzed the strength of disorder on the localization behavior \cite{Boguslawski}, as well, the absolute value still needs to be identified and associated with the factor $kl^*$ \cite{Ioffe}, including the mean free path length $l^*$ assumed to be in the range of $\sigma$. 
This can be done by analysis of the \cbs{} cone provided that a good signal to noise ratio is given. \cite{Akkermans}
However, it is not clear so far if the shape of the cone in our case is fully consistent with the suggested models.
Additionally, we identified a remarkable influence on the mean output spectrum by the shape of the refractive index power spectrum with upper limiting frequency $\kt{}$ representing an artificial, or effective, mobility edge.
Can its existence possibly mimic the transition in three dimensions?

All in all, subsequent experiments and analytics are necessary to characterize the presented system and to further prove its excellence for getting deeper insights into weak and strong localization phenomena.
In particular, it would be interesting to find a way of breaking the time reversal symmetry by dephasing pulses to quantitatively identify the influence of coherence in our observations. \cite{Micklitz}
Another phenomenon whose experimental implementation is highly on demand is coherent forward scattering of light \cite{Karpiuk, Ghosh} where our approach could be a candidate to provide the proper conditions.

%
\newpage

\begin{thebibliography}{99}
%
\bibitem{Bergmann} G.~Bergmann, ``Weak localization in thin films: a time-of-flight experiment with conduction electrons,'' Phys. Rep. \textbf{107}, 1 (1984).
%
\bibitem{Kirkpatrick} T.~R.~Kirkpatrick, ``Localization of acoustic waves,'' Phys. Rev. B \textbf{31}, 5746 (1985).
%
\bibitem{Bayer} G.~Bayer and T.~Niederdr\"ank, ``Weak localization of acoustic waves in strongly scattering media,'' Phys. Rev. Lett. \textbf{70}, 3884 (1993).
%
\bibitem{Tourin} A.~Tourin, A.~Derode,P.~Roux, B.~A.~van~Tiggelen, and M.~Fink, ``Time-Dependent Coherent Backscattering of Acoustic Waves,'' Phys. Rev. Lett. \textbf{79}, 3637 (1997).
%
\bibitem{Larose} E.~Larose, L.~Margerin, B.~A.~van~Tiggelen, and M.~Campillo, ``Weak Localization of Seismic Waves,'' Phys. Rev. Lett. \textbf{93}, 048501 (2004).
%
\bibitem{Stellmach} Ch.~Stellmach, H.~Abele, A.~Boucher, D.~Dubbers, U.~Schmidt, and P.~Geltenbort, ``On the Anderson-localization of ultra-cold neutrons,'' Nucl. Instr. and Meth. A \textbf{440}, 744 (2000).
%
\bibitem{Jendrzejewski} F.~Jendrzejewski, K.~M\"uller, J.~Richard, A.~Date, T.~Plisson, P.~Bouyer, A.~Aspect, A., and V.~Josse, ``Coherent Backscattering of Ultracold Atoms,'' Phys. Rev. Lett. \textbf{109}, 195302 (2012).
%
\bibitem{vAlbada} M.~P.~van~Albada, and A.~Lagendijk, ``Observation of Weak Localization of Light in a Random Medium,'' Phys. Rev. Lett. \textbf{55}, 2692 (1985).
%
\bibitem{Wolf} P.-E.~Wolf and G.~Maret, ``Weak Localization and Coherent Backscattering of Photons in Disordered Media,'' Phys. Rev. Lett. \textbf{55}, 2696 (1985).
%
\bibitem{Labeyrie1999} G.~Labeyrie, F.~de~Tomasi, J.-C.~Bernard, C.~A.~M\"uller, C.~Miniatura, and R.~Kaiser, ``Coherent Backscattering of Light by Cold Atoms,'' Phys. Rev. Lett. \textbf{83}, 5266 (1999).
%
\bibitem{Bidel} Y.~Bidel, B.~Klappauf, J.~C.~Bernard, D.~Delande, G.~Labeyrie, C.~Miniatura, D.~Wilkowski, and R.~Kaiser, ``Coherent light transport in a cold strontium cloud,'' Phys. Rev. Lett. \textbf{88}, 203902 (2002).
%
\bibitem{Wiersma} D.~S.~Wiersma, M.~P.~van~Albada, and A.~Lagendijk, `` An accurate technique to record the angular distribution of backscattered light,'' Rev. Sci. Instr. \textbf{66}, 5473 (1995).
%
\bibitem{Gross} P.~Gross, M.~St\"orzer, S.~Fiebig, M.~Clausen, G.~Maret, and C.~M.~Aegerter, ``A precisemethod to determine the angular distribution of backscattered light to high angles,'' Rev. Sci. Instr. \textbf{78}, 033105 (2007).
%
\bibitem{Aegerter} C.~Aegerter and G.~Maret, ``Coherent Backscattering and Anderson Localization of Light,'' Prog. Optics \textbf{52,} 1 (2009).
%
\bibitem{Skipetrov2004} S.~E.~Skipetrov and B.~A.~van Tiggelen, ``Dynamics of weakly localized waves,'' Phys. Rev. Lett. \textbf{92,} 113901 (2004).
%
\bibitem{Skipetrov2006} S.~E.~Skipetrov and B.~A.~van Tiggelen, ``Dynamics of Anderson localization in open 3D media,'' Phys. Rev. Lett. \textbf{96,} 043902 (2006).
%
\bibitem{Abrahams} E.~Abrahams, P.~W.~Anderson, D.~C.~Licciardello, and T.~V.~Ramakrishnan, ``Scaling Theory of Localization: Absence of Quantum Diffusion in Two Dimensions,'' Phys. Rev. Lett. \textbf{42,} 673 (1979).
%
\bibitem{Berkovits} R.~Berkovits and M.~Kaveh, ``Propagation of waves through a slab near the Anderson transition: A local scaling approach,'' J. Phys.: Condens. Mater. \textbf{2,} 307 (1990).
%
\bibitem{vTiggelen} B.~A.~van~Tiggelen, A.~Lagendijk, A.~Tip, and G.~F.~Reiter,``Effect of resonant scattering on localization of waves,'' Europhys. Lett. \textbf{15,} 535 (1991).
%
\bibitem{Sperling} T.~Sperling, W.~Bührer, C.~M.~Aegerter, and G.~Maret, ``Direct determination of the transition to localization of light in three dimensions,'' Nat. Photon. \textbf{7,} 48 (2013).
%
\bibitem{Labeyrie2012} G.~Labeyrie, T.~Karpiuk, J.-F.~Schaff, B.~Gremaud, C.~Miniatura, and D.~Delande, ``Enhanced backscattering of a dilute Bose-Einstein condensate,'' Europhys. Lett. \textbf{100,} 66001 (2012).
%
\bibitem{Anderson} P.~W.~Anderson, ``Absence of Diffusion in Certain Random Lattices,'' Phys. Rev. \textbf{109,} 1492 (1958).
%
\bibitem{Schwartz} T.~Schwartz, G.~Bartal, S.~Fishman, and M.~Segev, ``Transport and Anderson localization in disordered two-dimensional photonic lattices,'' Nature \textbf{446,} 52 (2007).
%
\bibitem{Levi2011} L.~Levi, M.~Rechtsman, B.~Freedman, T.~Schwartz, O.~Manela, and M.~Segev, ``Disorder-Enhanced Transport in Photonic Quasicrystals,'' Science \textbf{332,} 1541 (2011).
%
\bibitem{Levi2012} L.~Levi, Y.~Krivolapov, S.~Fishman, and M.~Segev, ``Hyper-transport of light and stochastic acceleration by evolving disorder,'' Nat. Phys. \textbf{8,} 912 (2012).
%
\bibitem{Boguslawski} M.~Boguslawski, S.~Brake, J.~Armijo, F.~Diebel, P.~Rose, and C.~Denz, ``Analysis of transverse Anderson localization in refractive index structures with customized random potential,'' Opt. Express \textbf{21,} 31713 (2013).
%
\bibitem{Verma} M.~Verma, D.~K.~Singh, P.~Senthilkumaran, J.~Joseph, and H.~C.~Kandpal, ``Ultrasensitive and fast detection of denaturation of milk by Coherent backscattering of light,'' Sci. Rep. \textbf{4,} 7257 (2014).
%
\bibitem{Cottrell} D.~M.~Cottrell, J.~M.~Craven, and J.~A.~Davis, ``Nondiffracting random intensity patterns,'' Opt. Lett. \textbf{32,} 298 (2007).
%
\bibitem{Durnin} J.~Durnin, ``Exact solutions for nondiffrating beams. I. The scalar theory,'' J. Opt. Soc. Am. A \textbf{4,} 651 (1987).
%
\bibitem{Weilnau} C.~Weilnau, M.~Ahles, J.~Petter, D.~Tr\"ager, J.~Schr\"oder, and C.~Denz, ``Spatial optical (2+1)-dimensional scalar- and vector-solitons in saturable nonlinear media,'' Ann. Phys. \textbf{11,} 573 (2002).
%
\bibitem{Hall} T.~J.~Hall, R.~Jaura, L.~M.~Connors, and P.~D.~Foote, ``The photorefractive effect--a review,'' Prog. Quant. Electron. \textbf{10,} 77 (1985).
%
\bibitem{Terhalle} B.~Terhalle, D.~Tr\"ager, L.~Tang, J.~Imbrock, and C.~Denz, ``Structure analysis of two-dimensional nonlinear self-trapped photonic lattices in anisotropic photorefractive media,'' Phys. Rev. E \textbf{74,} 057601 (2006).
%
\bibitem{Sanchez} L.~Sanchez-Palencia, D.~Clement, P.~Lugan, P.~Bouyer, G.~V.~Shlyapnikov, and A.~Aspect, ``Anderson Localization of Expanding Bose-Einstein Condensates in Random Potentials,'' Phys. Rev. Lett. \textbf{98,} 210401 (2007).
%
\bibitem{Gurevich} E.~Gurevich and O.~Kenneth, ``Lyapunov exponent for the laser speckle potential: A weak disorder expansion,'' Phys. Rev. A \textbf{79,} 063617 (2009).
%
\bibitem{Ioffe} A.~F.~Ioffe and A.~R.~Regel, ``Non-crystalline, amorphous and liquid electronic semiconductors,'' Progr. Semicond. \textbf{4,} 237 (1960).
%
\bibitem{Akkermans} E.~Akkermans, P.~E.~Wolf, and R.~Maynard, ``Coherent backscattering of light by disordered media: Analysis of the peak line shape,'' Phys. Rev. Lett. \textbf{56,} 1471 (1986).
%
\bibitem{Micklitz} T.~Micklitz, C.~A.~M\"uller, and A.~Altland, ``Echo spectroscopy of Anderson localization,'' arXiv:1406.6915 (2014).
%
\bibitem{Karpiuk} T.~Karpiuk, N.~Cherroret, K.~L.~Lee, B.~Gr\'emaud, C.~A.~M\"uller, and C.~Miniatura, ``Coherent Forward Scattering Peak Induced by Anderson Localization,'' Phys. Rev. Lett. \textbf{109,} 190601 (2012).
%
\bibitem{Ghosh} S.~Ghosh, N.~Cherroret, B.~Gremaud, C.~Miniatura, and D.~Delande, ``Coherent forward scattering in two-dimensional disordered systems,'' Phys. Rev. A \textbf{90,} 063602 (2014).
\end{thebibliography}
\end{document}